\documentclass[lettersize,onecolumn]{IEEEtran}
\usepackage{amsmath,amsfonts}
\usepackage{algorithmic}
\usepackage{algorithm}
\usepackage{array}
\usepackage[caption=false,font=normalsize,labelfont=sf,textfont=sf]{subfig}
\usepackage{textcomp}
\usepackage{stfloats}
\usepackage{url}
\usepackage{verbatim}
\usepackage{graphicx}
\usepackage{cite}
\makeatletter
\def\@cite#1#2{\textsuperscript{[#1\if@tempswa , #2\fi]}}
\makeatother
\usepackage{amssymb}
\usepackage{xcolor}

\begin{document}

\title{Distributed Coherent Optical Computing\\ via Injection-Locked Photonic Networks}

\author{Shenghan Gao,~\IEEEmembership{Student Member,~IEEE}, Kathy Lüdge,~\IEEEmembership{Member,~IEEE}, Francesco Da Ros,~\IEEEmembership{Senior Member,~IEEE},\\ Nathan Youngblood,~\IEEEmembership{Member,~IEEE}
\thanks{This work was supported in part by the Office of Naval Research (ONR N00014-26-1-2192) and Air Force Office of Scientific Research (AFOSR FA9550-24-1-0064 Young Investigator Award)}%
\thanks{S. Gao and N. Youngblood are with the Department of Electrical and Computer Engineering, Swanson School of Engineering, University of Pittsburgh, Pittsburgh, PA 15261 USA.}%
\thanks{F. Da Ros is with DTU Electro, Technical University of Denmark, 2800 Kgs. Lyngby, Denmark}%
\thanks{K. Lüdge is with the Institute of Physics, Ilmenau University of Technology, 98693 Ilmenau, Germany.}
}



\maketitle

\begin{abstract}
    Coherent photonic computing uses both the phase and amplitude of light to implement linear operations such as dot products and matrix multiplication but requires phase stability between the interfering paths. This poses a challenge for such strategies when optical data is generated at a remote source due to environmental phase variations in fiber. Conventional approaches to distributed computing rely on optical-to-electrical conversion and buffering, limiting truly real-time and distributed computation. Here, we propose a new strategy via optical injection locking to enable distributed, real-time coherent optical processing without unnecessary conversions in the optical-to-electrical or analog-to-digital domains. Using a semiconductor laser rate-equation model, we explore the conditions required for stable operation by sweeping the power injection ratio, frequency detuning, and modulation conditions of the remote and injected lasers. Our results indicate that higher injection powers broaden the locking margin but more readily exhibit frequency-selective features associated with relaxation oscillations and increased amplitude-phase mixing, whereas lower injection powers yield a narrower, but more predictable operating window which remains stable under large modulation depth. End-to-end symbol-sequence simulations with balanced detection and temporal integration further confirm that reducing the injection ratio suppresses residual remote-modulation components in the injected laser output and improves computational accuracy. Overall, our study provides guidance and design trade-offs for remote coherent detection and distributed coherent photonic computing enabled by injection locking.
\end{abstract}


\section{Introduction}

Over the past decades, neural networks and data-centric workloads have continued to increase the demand for computation, while improvements in general-purpose digital hardware face growing throughput and energy-efficiency constraints. Across many applications, large-scale linear-algebra primitives---such as dot products and matrix multiplications---remain central building blocks and are often accompanied by substantial data-movement and memory overhead. Against this backdrop, there is increasing interest in alternative physical implementations that could execute selected linear operators with higher throughput and potentially better energy efficiency, thereby alleviating pressure on purely digital processing.

\begin{figure}[H]
  \centering
  \includegraphics[width=0.9\linewidth]{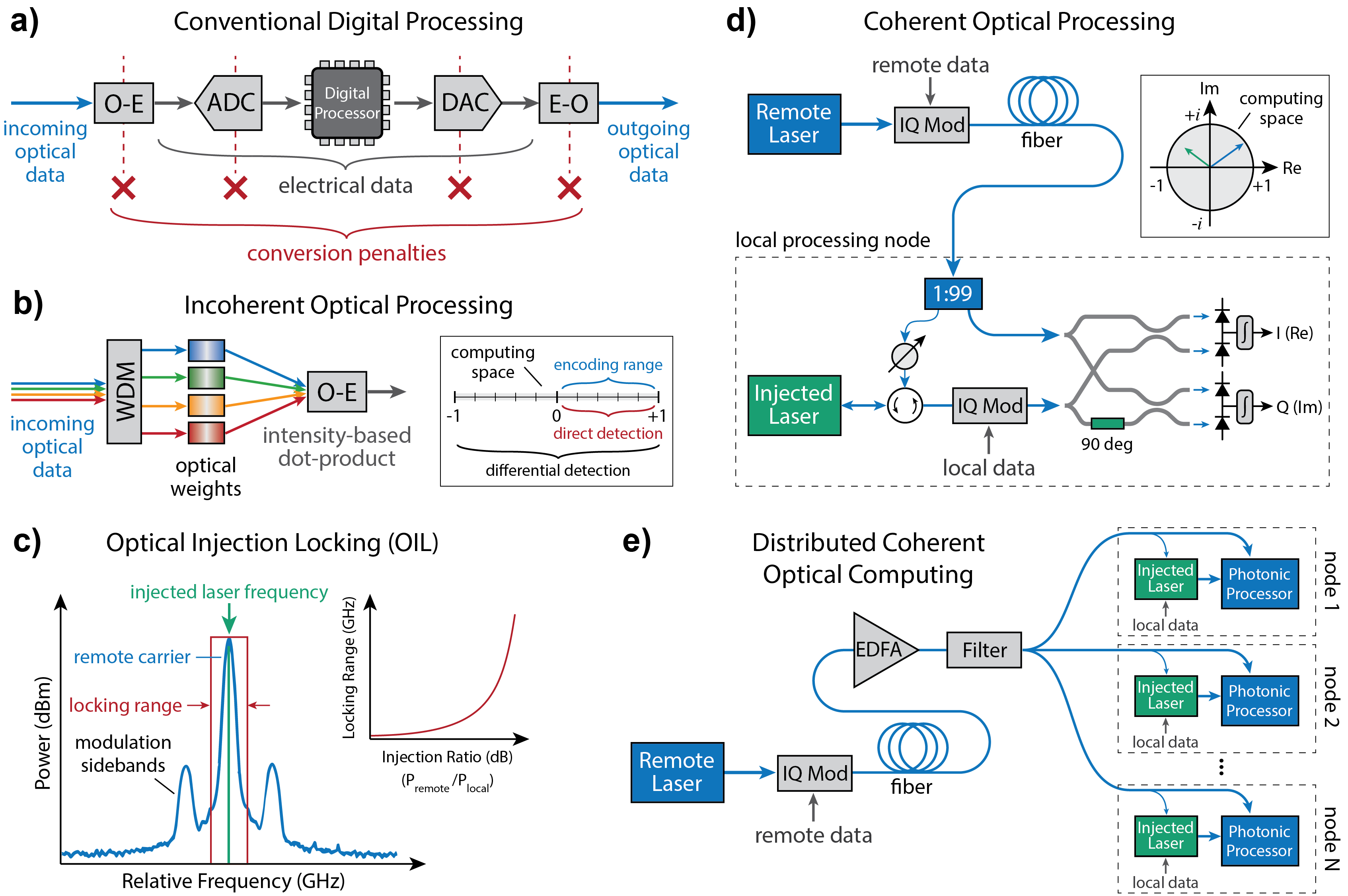}
  \caption{\textbf{Overview of distributed computing strategies and proposed architecture. a)} Conventional digital computing approaches require multiple conversion penalties at O-E and A-D conversion boundaries when networking distributed computing nodes. \textbf{b)} Incoherent optical architectures are able to perform real-time processing of data from a remote source, but encoding is limited to positive real-valued numbers and by the number of independent channels (wavelength, mode, polarization, etc.) resulting in limited problem dimension. \textbf{c)} OIL allows a coherent copy of a remote source laser while filtering out frequency components outside the locking range. Locking range can be tuned through the injection power of the remote laser relative to the injected laser (see inset). \textbf{d)} Proposed platform for distributed, coherent optical computing. Data from a remote laser is stripped from the carrier frequency via OIL, creating a coherent, unmodulated copy of the remote laser. Subsequent modulation and interference allow real-time, coherent processing directly on the remote signal. \textbf{e)} Extending OIL concept to multiple nodes for scalable networking of distributed photonic computing chips. Each node independently locks to the optical carrier of original remote laser, enabling independent parallel computations to be performed on the original signal.}
  \label{fig:fig1}
\end{figure}

Although photonics has traditionally enabled high-speed interconnects and data transport, significant effort has been dedicated to developing programmable photonic circuits as a substrate for physically mapping linear operations to the optical domain, leveraging optical bandwidth and parallelism to reduce the amount of work that must be handled digitally \cite{Bogaerts2020Nature_ProgrammablePICs_Review,Tait2014JLT_BroadcastAndWeight,Feldmann2021Nature_PhotonicTensorCore,Tait2017SciRep_NeuromorphicWeightBanks,Shen2017NatPhoton_CoherentNanophotonicDL,Hamerly2019PRX_PhotoelectricMultiplication}. This concept is particularly appealing when considering the growing significance that optical interconnects play in today's datacenters to network digital processors. While optical interconnects allow digital processors to communicate at much higher bandwidths than copper wire, this communication is dominated by penalties stemming from conversions between optical-to-electrical (O-E) and analog-to-digital (A-D) domains as illustrated in Fig.~\ref{fig:fig1}a. 

In an ideal world, analog optical data received from a remote processor would be processed in the optical domain in real-time, thus removing energy penalties from domain conversions and latency penalties from digital buffering. While real-time processing of optical data from a remote source has been demonstrated in certain architectures \cite{Xu2021Nature_11TOPS_ConvAccelerator,Hamerly2024JLT_Netcast,Sludds2022Science_NetcastEdge}, this approach relies on incoherent optical signaling where information from the remote source is encoded on the amplitude and wavelength of light. This limits the encoding range to positive, real-valued numbers between 0 and $+1$ (Fig.~\ref{fig:fig1}b). Differential detection can expand the output range to at most real-valued numbers ranging between $-1$ and $+1$ through various routing and balanced detection approaches \cite{Tait2017SciRep_NeuromorphicWeightBanks,Tait2014JLT_BroadcastAndWeight}.

Coherent architectures, in contrast, preserve phase information such that optical interference produces a cross term directly related to the overlap of two complex fields \cite{Shen2017NatPhoton_CoherentNanophotonicDL}. This property lends itself naturally to dot-product and multiply-accumulate primitives and even supports complex-valued numbers through I/Q or amplitude-phase encoding \cite{Shen2017NatPhoton_CoherentNanophotonicDL,Zhou2025_QA_PhotonicHomodyne,RahimiKari2024Optica_CoherentPlatform,Chen2023NatPhoton_VCSEL,Zhang2021NatCommun_ComplexONC}. A challenge for such approaches is the requirement of mutual coherence---the two interfering paths must share a coherent reference such that their relative phase remains sufficiently stable over the integration time of the detector \cite{Shen2017NatPhoton_CoherentNanophotonicDL}. For example, in prior work by Chen et al. \cite{Chen2023NatPhoton_VCSEL}, OIL by a continuous-wave (CW) lead laser was used to establish a shared carrier frequency and phase reference among an array of independent VCSELs to support coherent interference-based readout in a neural-network setting. However, in this demonstration (as well as others), optical data for both matrix weights and input vectors was local to the VCSEL array. This need for chip-level data locality in coherent architectures stems directly from practical challenges in maintaining phase coherence across arbitrary distances and places an upper bound on the efficiency and latency of optically networked clusters of coherent photonic computing chips if conventional O-E / A-D conversions and buffering are used to network such chips.



To address these challenges, this paper proposes optical injection locking (OIL) as a mechanism for maintaining coherence across distributed optical nodes. We leverage a technique previously demonstrated in coherent optical communication where OIL is used to regenerate a local oscillator (LO) directly from the modulated optical carrier \cite{LiuSlavik2020OILTutorial,Liu2015JLT_HomodyneOFDM_OIL_CarrierRecovery}. Here, a fraction of the \emph{remote} laser is injected into an \emph{injected} laser which then locks to the phase and frequency of the remote laser, provided the frequency detuning between the two lasers is within the locking range \cite{Lang1982InjectionLockingProperties,Mogensen1985JQE_LockingStability_SemiconductorInjection}. As illustrated in Fig.~\ref{fig:fig1}c, the locking range depends on the injection ratio and laser parameters, providing a tunable operating window for phase synchronization. Crucially, if the modulation sidebands carried by the remote laser fall outside the locking range bandwidth, they will be filtered out, and the injected laser will simply be a direct copy of the unmodulated remote laser \cite{Liu2015JLT_HomodyneOFDM_OIL_CarrierRecovery}.

We propose applying this technique to coherent optical processing \cite{Zhou2025_QA_PhotonicHomodyne} where analog data from a remote source (e.g., complex vector $\bf \tilde{x}$) is filtered from the carrier via OIL, generating a coherent copy of the unmodulated source. The injected laser is then remodulated with local data (e.g., complex vector $\bf \tilde{y}$), interfered with the original optical signal, and temporally integrated to yield the complex dot-product ${\bf\tilde{x}} \cdot {\bf\tilde{y}} = \tilde{z}$ as illustrated in Fig.~\ref{fig:fig1}d. This approach provides a powerful scaling advantage as a single remote source can be fanned out to multiple nodes, where independent injected lasers can be injection-locked to the same remote source carrier reference \cite{LiuSlavik2020OILTutorial}, providing a consistent phase reference across distributed processors and enabling parallel operation (Fig.~\ref{fig:fig1}e). Notably, this approach allows independent tracking of phase and frequency variations across the network without the need for multiple active control loops, provided polarization is maintained and phase variations are within the bandwidth of the locking range.

The objective of this work is to quantitatively assess key design trade-offs of this approach using a rate-equation simulation framework for semiconductor lasers \cite{Lang1982InjectionLockingProperties,WieczorekKrauskopfLenstra1999Bifurcations}. In Section II, We begin by mapping the locking range as a function of detuning and injection ratio to determine the operating region and filtering bandwidth of the injected laser. In Section III,  we analyze stability and modulation-transfer behavior across injection regimes, modulation frequency, and semiconductor linewidth enhancement factor $\alpha$, with particular attention to their influence on relaxation-oscillation (RO) dynamics in the injected laser \cite{Lau2008JQE_FreqRespEnh_OIL,WieczorekKrauskopfLenstra1999Bifurcations}. In Section IV, We quantify how realistic symbol sequences impact dot-product accuracy at different injection ratios. These results aim to clarify practical design parameters that are more likely to support stable operation and to provide theoretical guidance for remote coherent detection and distributed coherent optical computing architectures.


\section{Rate-Equation Model and Locking-Region Mapping}

\subsection{From Adler Locking Range to Semiconductor-Laser Injection Locking}

Injection locking can be understood from a classical oscillator synchronization viewpoint. Under a small-signal approximation that focuses on the relative phase dynamics, Adler derived a closed-form locking condition for an injection-locked oscillator~\cite{Adler1946ProcIRE_Locking} where the frequency detuning $\Delta\omega_0$ between the two oscillators must satisfy the explicit locking-range inequality:
\begin{equation}
\left|\Delta\omega_0\right|
\le \Delta\omega_{\max}
= \frac{\omega_0}{2Q}\frac{E_{in}}{E_0},
\label{eq:adler_locking_range}
\end{equation}
here, $\omega_0$ is the oscillation angular frequency, $Q$ is the quality factor, $E_{in}$ is the amplitude of the input field, and $E_0$ is the free-running steady-state oscillation amplitude of the injected oscillator. Equation~\eqref{eq:adler_locking_range} directly indicates that the locking range increases with injection strength and decreases with the injected oscillator's quality factor.

For semiconductor lasers, the concept of a frequency-dependent locking interval remains applicable, but the locking boundary is strongly influenced by amplitude--phase coupling. This means that carrier perturbations modulate both gain and refractive index, leading to a non-zero linewidth enhancement factor $\alpha$ which couples amplitude dynamics into phase dynamics. Consequently, the locking range is generally asymmetric with respect to detuning and its stability depends on semiconductor-specific carrier dynamics \cite{Lang1982InjectionLockingProperties,HenryOlssonDutta1985LockingRangeStability,Mogensen1985JQE_LockingStability_SemiconductorInjection}. This can be summarized by the unified expression \cite{LiuSlavik2020OILTutorial}:

\begin{equation}
-\kappa \sqrt{1+\alpha^2}\sqrt{\frac{P_{\mathrm{remote}}}{P_{\mathrm{0}}}}
= \Delta\omega_{\min}
< \Delta\omega
< \Delta\omega_{\max}
= \kappa\sqrt{\frac{P_{\mathrm{remote}}}{P_{\mathrm{0}}}},
\label{eq:semiconductor_locking_range}
\end{equation}
where $P_{\mathrm{remote}}/P_{\mathrm{0}}$ is the injection ratio (remote laser power over free-running output power of the injected laser), $\kappa$ is an effective coupling coefficient, and $\alpha$ explicitly enlarges the negative-detuning boundary through $\sqrt{1+\alpha^2}$. While Eqs.~\eqref{eq:adler_locking_range}--\eqref{eq:semiconductor_locking_range} provide direct intuition on how the locking range scales with injection strength, they primarily characterize the existence of locked solutions. In practice, intensity and carrier dynamics (e.g., relaxation oscillations) also shape the stability and transient behavior of locked operation.

Thus, to better understand the locking region and dynamical behavior of injection locking when a semiconductor laser serves as the injected laser, we simulate the complex, dynamic evolution of the injected laser's complex field, carrier number, and gain based on well-established laser rate equations. This allows us to accurately describe the time evolution of a semiconductor laser's intracavity complex field and carrier population, including injection-locking dynamics \cite{Fragkos2012JLT_AmpLimiter_OIL,Dong2024RC_QDLaser,Muhlnickel2024CommunPhys_SpinVCSEL_RC,Pausch2012NJP_QDInjectionDynamics}. Using this framework, we simulate multiple injection-locking scenarios over broad injection ratios and detuning ranges.

\subsection{Laser Rate Equations and 2D Locking-Region Mapping}

We use a Lang--Kobayashi-style semiconductor-laser rate-equation model consistent with the formulation employed by Fragkos \emph{et al.}~\cite{Fragkos2012JLT_AmpLimiter_OIL}. The injected laser is described by the slowly varying complex intracavity field $E(t)$ and carrier number $N(t)$:
\begin{align}
\frac{dE}{dt} &=
\frac{1}{2}\left(1+i\alpha\right)G\!\left(N,|E|^2\right)E
+ \kappa\,E_{\mathrm{inj}}(t),
\label{eq:rateE}\\
\frac{dN}{dt} &=
R_p - \frac{N}{\tau_n}
- g\!\left(N,|E|^2\right)|E|^2,
\label{eq:rateN}
\end{align}
where $\alpha$ is the linewidth enhancement factor, $\tau_n$ is the carrier lifetime, $R_p$ is the pump (bias) term, and $\kappa$ is the injection coupling coefficient. We model the saturated gain as:
\begin{align}
g\!\left(N,|E|^2\right)
&=\frac{g_0\left(N-N_0\right)}{1+s|E|^2},
\label{eq:satgain}\\
G\!\left(N,|E|^2\right)
&= g\!\left(N,|E|^2\right) - \frac{1}{\tau_s},
\label{eq:netgain}
\end{align}
where $g_0$ is the differential gain parameter, $N_0$ is the transparency carrier number, $s$ is the gain saturation factor, and $\tau_s$ is the photon lifetime. The injection term $E_{\mathrm{inj}}(t)$ represents a fraction of the remote-laser field coupled into the injected laser. In this section, we consider an unmodulated injection field to map the carrier-level locking region. Modulation-dependent effects are analyzed in later sections.

\begin{figure}[H]
  \centering
  \includegraphics[width=0.9\linewidth]{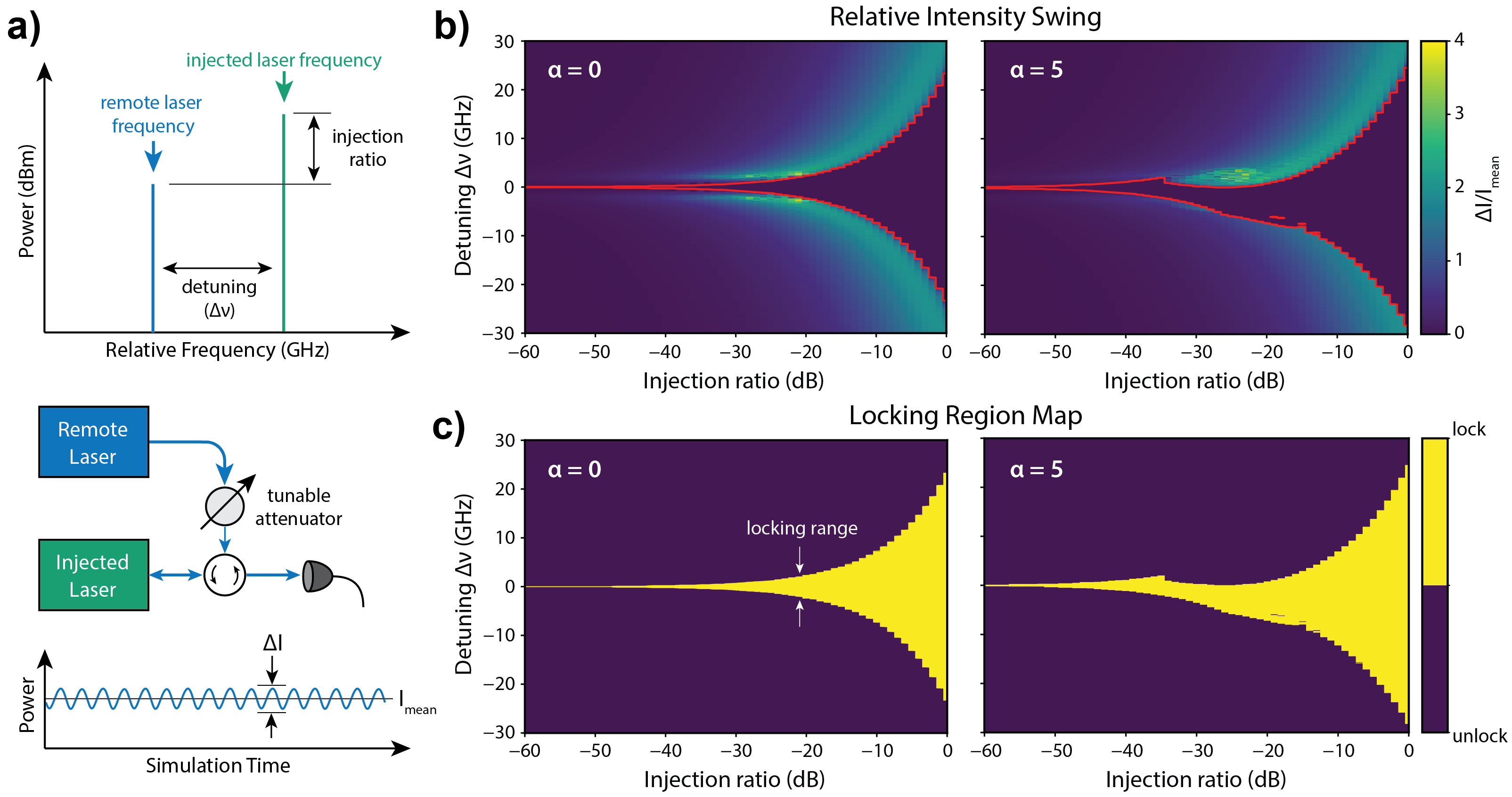}
  \caption{\textbf{Mapping locking range for various detuning and injection ratios. a)} Conceptual overview of parameter sweep and simulated analysis used to extract locking region from laser rate equations. \textbf{b)} Relative intensity swing of injected laser under varying injection ratios and detunings for $\alpha=0$ and $\alpha=5$. Boundaries where $\Delta I/\bar{I}\le 0.10$ is indicated in red. \textbf{c)} Two-dimensional locking-region maps for $\alpha=0$ and $\alpha=5$ considering both phase and amplitude stability. The locked region is identified using the steady-state criteria $\Delta I/\bar{I}\le 0.10$ and $\Delta\phi\le 0.10~\mathrm{rad}$.}
  \label{fig:fig2}
\end{figure}

Internal parameters are selected to match those reported in~\cite{Fragkos2012JLT_AmpLimiter_OIL}, providing a consistent physical baseline for our simulations. Table~\ref{tab:fragkos_params} summarizes the values used in our modeling. To obtain reproducible, deterministic locking boundaries in our OIL stability maps, we adopt the following assumptions: (i) spontaneous-emission noise is not included in the locking-region scans; (ii) the field is represented in the remote laser rotating frame (implemented via real/imaginary components in the numerical solver); (iii) the injection ratio is defined as $R_{\mathrm{inj}}=P_{\mathrm{remote}}/P_0$ (in dB), where $P_0$ is the free-running steady-state intensity of the injected laser computed analytically from the same model; and (iv) the pump bias is fixed at $1.15\times$ threshold in all simulations.

\begin{table}[H]
\centering
\caption{Simulation parameters taken from \cite{Fragkos2012JLT_AmpLimiter_OIL}.}
\label{tab:fragkos_params}
\begin{tabular}{l c c}
\hline
Parameter & Symbol & Value \\
\hline
Differential gain (ps$^{-1}$) & $g_0$ & $1\times 10^{-8}$ ps$^{-1}$ \\
Gain saturation factor & $s$ & $5\times 10^{-7}$ \\
Linewidth enhancement factor & $\alpha$ & $0,\ 5$ \\
Carrier lifetime & $\tau_n$ & $2$ ns \\
Transparency carrier number & $N_0$ & $2.1\times 10^8$ \\
Round-trip time & $t_{\mathrm{rt}}$ & $7$ ps \\
Photon lifetime & $\tau_s$ & $3.2$ ps \\
Cavity rate & $\gamma_c$ & $1/t_{\mathrm{rt}}$ \\
Injection coupling & $\kappa$ & $\kappa_{\mathrm{scale}}\cdot \gamma_c$ \\
Pump bias & $R_p$ & $1.15\times R_{\mathrm{th}}$ \\
\hline
\end{tabular}
\end{table}

As shown in Fig.~\ref{fig:fig2}a, we map the locking region by scanning a grid over injection ratio and detuning, and integrating Eqs.~\eqref{eq:rateE}--\eqref{eq:rateN} using a fixed-step fourth-order Runge--Kutta method~\cite{Dong2024RC_QDLaser}. For each grid point $(R_{\mathrm{inj}},\Delta\nu)$, trajectories are simulated over a sufficiently long window, and only the tail segment is used to evaluate steady-state metrics. We quantify locking using two complementary measures:

\begin{description}
    \item[(i) relative intensity swing:] 
    \begin{equation}
    \frac{\Delta I}{\bar I} \triangleq \frac{I_{\max}-I_{\min}}{I_{\mathrm{mean}}},
    \quad I(t)=|E(t)|^2,
    \label{eq:dI}
    \end{equation}

    \item[(ii) phase excursion:]
    \begin{equation}
    \Delta\phi \triangleq \max(\phi_{\mathrm{unwrap}})-\min(\phi_{\mathrm{unwrap}}),
    \label{eq:dphi}
    \end{equation}
\end{description}
where $\phi(t)=\arg(E(t))$ is unwrapped over the analysis window to avoid $2\pi$ discontinuities. A grid point is labeled as locked if it satisfies $\Delta I/\bar I \le 0.10$ and $\Delta\phi \le 0.10~\mathrm{rad}$. The resulting locking region maps are shown in Fig.~\ref{fig:fig2} for $\alpha=0$ and $\alpha=5$. In Fig.~\ref{fig:fig2}b, the maps visualize the relative intensity swing $\Delta I/\bar{I}$ versus injection ratio $(R_{\mathrm{inj}})$ and relative detuning between the remote and injected lasers $(\Delta\nu)$. The red-outlined region indicates the boundary of stability where $\Delta I/\bar{I}\le 0.10$. Fig.~\ref{fig:fig2}c combines the intensity and phase criteria to identify the locking range: the highlighted region corresponds to points that satisfy both $\Delta I/\bar{I}\le 0.10$ and $\Delta\phi\le 0.10~\mathrm{rad}$ (see Methods Section for more information).

Two trends are apparent from these results. First, for both $\alpha=0$ and $\alpha=5$, increasing the injection ratio generally broadens the locking range, consistent with the expected dependence of locking margin on injection strength described in Eq. \eqref{eq:semiconductor_locking_range}. Second, for $\alpha=5$, the locking region becomes increasingly asymmetric as injection ratio increases, reflecting the role of amplitude--phase coupling in shifting the negative-frequency detuning boundary. Additionally, at higher injection ratios, the boundary and interior of the mapped region can exhibit localized irregular regions of instability, suggesting increased dynamical sensitivity and the onset of more complex nonlinear behavior under strong optical injection. Here we treat these features as qualitative indicators motivating further analysis, consistent with prior studies showing that optically injected semiconductor lasers can exhibit rich bifurcation structures and frequency-dependent nonlinear dynamics \cite{WieczorekKrauskopfLenstra1999Bifurcations,Mogensen1985JQE_LockingStability_SemiconductorInjection,Lau2008JQE_FreqRespEnh_OIL}.


\section{Frequency-Dependent Transfer Functions under Injection Locking}
In the distributed coherent computing approach considered in this work, the signal injected from the remote source is \emph{modulated} rather than a pure continuous-wave carrier. Therefore, in this section, we explicitly use a modulated remote laser for optical injection in our simulations and evaluate how the modulated signal propagates through the injection-locking dynamics (illustrated in Fig.~\ref{fig:fig3}a). Our desire is to minimize the transfer of this unwanted signal from the remote laser onto the injected laser while still coherently locking to the remote laser's optical carrier frequency. Therefore, a central metric we use is the frequency-dependent transfer function, defined as the ratio between the AC component extracted from the injected laser output and the corresponding AC component of the remote-laser injection signal at the same modulation frequency. This transfer-function view complements carrier-level locking checks by quantifying the degree of modulation coupled to the injected laser's output.

\subsection{Extracting AM/PM Transfer Functions from Rate-Equation Outputs}

In the following simulations, we inject a remote laser carrying either amplitude modulation (AM) or phase modulation (PM) into the Lang--Kobayashi-style rate-equation model introduced in Sec.~II~\cite{Fragkos2012JLT_AmpLimiter_OIL}. The injected field is written as:s
\begin{equation}
E_{\mathrm{inj}}(t) = A_m(t)\,e^{j\theta_m(t)},
\end{equation}
where $A_m(t)$ and $\theta_m(t)$ denote the instantaneous amplitude and phase, respectively. We consider two single-tone modulation formats at angular frequency $\Omega = 2\pi f_{\mathrm{mod}}$. For AM, the phase remains constant, while the amplitude is sinusoidally modulated. The opposite case is used for PM:

AM:
\begin{equation}
A_m(t) = E_{0} + \Delta E \sin(\Omega t), \ \ \ \ \ 
\theta_m(t)=\phi_0
\end{equation}

PM:
\begin{equation}
\theta_m(t) = \phi_0 + \Delta\phi \sin(\Omega t), \ \ \ \ \ 
A_m(t) = E_{0}
\end{equation}
The normalized modulation depths ($m_{a}=\Delta E/E_{0}$ and $m_\phi=\Delta\phi/\pi$) of the AM and PM signals are controlled through the remote modulation amplitudes $\Delta E$ and $\Delta\phi$. We then extract the resulting AC components at the modulation frequency from both the remote laser field and the injected laser response to define AM and PM transfer functions~\cite{LiuSlavik2020OILTutorial}. For a given $(\Delta\nu, R_{\mathrm{inj}}, f_{\mathrm{mod}})$, we compute the single-sided fundamental amplitude at $f_{\mathrm{mod}}$ for each channel using a Fast Fourier Transform (FFT) over a steady-state measurement window ($A_{\mathrm{a,s}}$ and $A_{\mathrm{\phi,s}}$). Further details on extracting these amplitudes and normalizing the signals can be found in the Appendix. The four small-signal transfer functions are then defined as
\begin{align}
H_{\mathrm{AM\to AM}} &= \frac{A_{\mathrm{a,s}}}{\Delta E}, &
H_{\mathrm{AM\to PM}} &= \frac{A_{\mathrm{\phi,s}}}{\Delta E}, \\
H_{\mathrm{PM\to PM}} &= \frac{A_{\mathrm{\phi,s}}}{\Delta\phi}, &
H_{\mathrm{PM\to AM}} &= \frac{A_{\mathrm{a,s}}}{\Delta\phi}.
\end{align}
where $A_{\mathrm{a,s}}$ and $A_{\mathrm{\phi,s}}$ are the extracted AM and PM frequency-dependent AC amplitudes from the injected laser. We report $|H|$ in dB (i.e., $20\log_{10}|H|$) and visualize the results as pseudo-color maps in subsequent parameter sweeps (injection-ratio sweep in Fig.~3, detuning sweep in Fig.~4, and modulation-depth sweep in Fig.~5).

Finally, we note that when the linewidth enhancement factor $\alpha \neq 0$, amplitude--phase coupling can induce crosstalk between AM and PM channels: AM modulation can generate a PM component in the injected laser response (AM$\to$PM), and PM modulation can generate an AM component (PM$\to$AM). Therefore, in addition to the direct channels (AM$\to$AM and PM$\to$PM), we explicitly track the transfer function between channels (AM$\to$PM and PM$\to$AM) as quantitative measures of modulation crosstalk under injection locking.

\subsection{Injection-Ratio Sweep: Transfer-Function Dependence and RO Peak Shift}

We first focus on the case where the remote laser and the injected laser are perfectly wavelength matched (fixed detuning at $\Delta\nu=0$) and sweep the injection ratio to quantitatively evaluate how the injected laser transmits or suppresses a modulation signal carried by the remote laser. To more clearly isolate small-signal modulation transfer associated with injection locking and to reduce the influence of sidebands and higher-order nonlinear effects induced by strong modulation, we use shallow modulation depths in the injected field: $m_a=0.1\%$ and $m_\phi=0.1\%$. For $\alpha=0$, amplitude--phase coupling is strongly suppressed and the cross-channel responses (AM$\to$PM and PM$\to$AM) are numerically negligible in this sweep; therefore, we primarily focus on the direct channels AM$\to$AM and PM$\to$PM~\cite{Lang1982InjectionLockingProperties,HenryOlssonDutta1985LockingRangeStability,Mogensen1985JQE_LockingStability_SemiconductorInjection}.

\begin{figure}[H]
  \centering
  \includegraphics[width=0.9\linewidth]{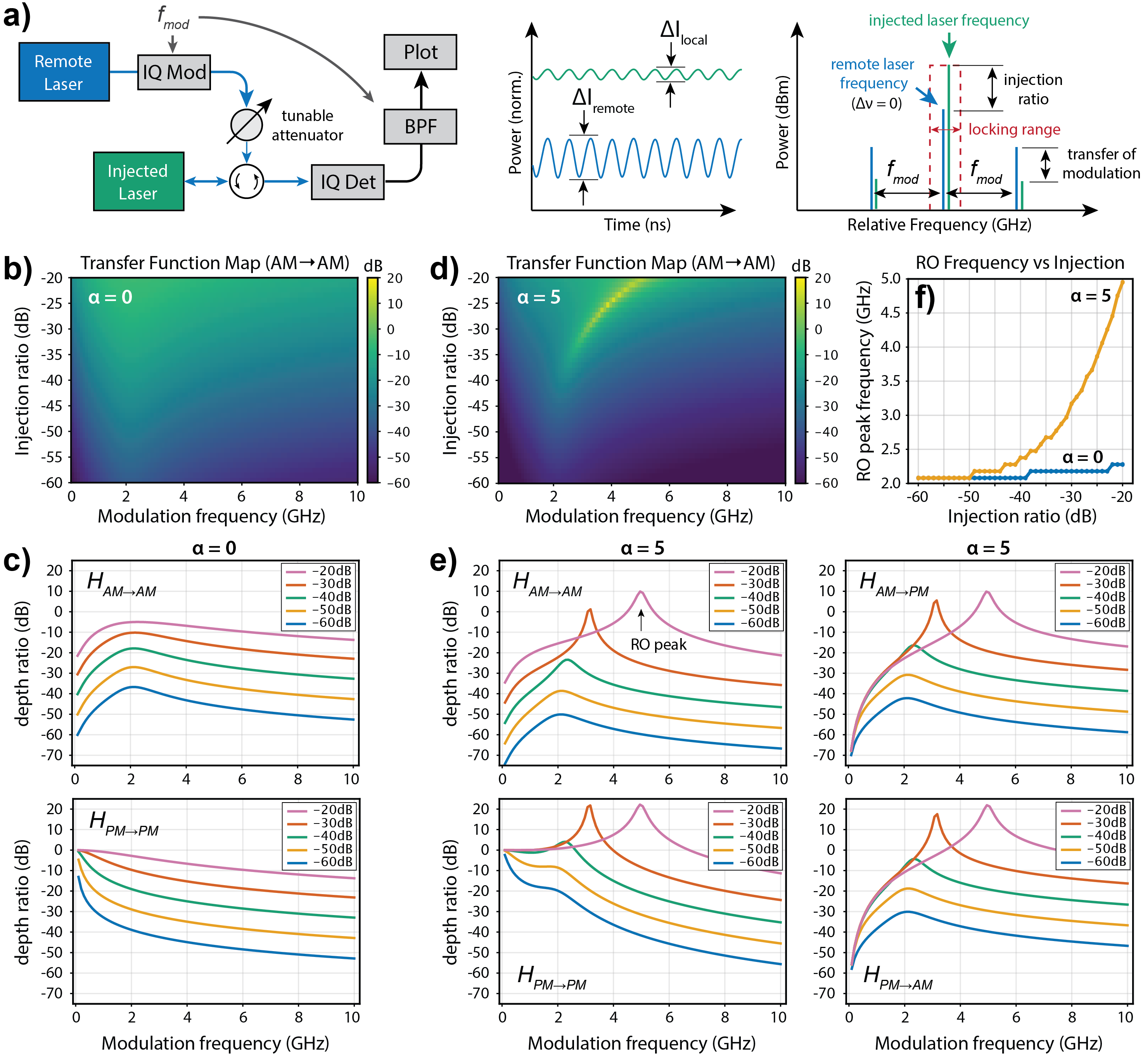}
  \caption{\textbf{Injection-ratio sweep at $\Delta\nu = 0$ with shallow modulation ($m_a = 0.1\%$, $m_\phi = 0.1\%$).} \textbf{a)} Illustration of simulation setup to extract frequency dependent transfer curves for amplitude (and phase) modulation. Panels \textbf{b)} and \textbf{d)} show AM$\to$AM transfer-function heatmaps over injection ratio ($-60$ to $-20$~dB) and $f_{\mathrm{mod}}$, with color indicating $20\log_{10}|H|$, for $\alpha = 0$ and $\alpha = 5$, respectively. \textbf{c)} AM$\to$AM and PM$\to$PM transfer function plots for $\alpha = 0$. \textbf{e)} Full transfer function plots for $\alpha = 5$, including AM$\to$AM and PM$\to$PM coupling as well as AM$\to$PM and PM$\to$AM inter-channel crosstalk. \textbf{f)} Extracted RO peak frequency versus injection ratio, showing a modest shift for $\alpha = 0$ ($\sim$2.0--2.2~GHz) and a larger shift for $\alpha = 5$ ($\sim$2.0--5~GHz).}
\label{fig:fig3}
\end{figure}

For each injection ratio from $-60$~dB to $-20$~dB, we compute the transfer magnitude $|H(f_{\mathrm{mod}})|$ over $f_{\mathrm{mod}}\in[0.1,10]$~GHz using 100 frequency points. Fig.~\ref{fig:fig3}c shows the AM$\to$AM and PM$\to$PM transfer functions for $\alpha=0$, while Fig.~\ref{fig:fig3}e shows the corresponding results for $\alpha=5$ and additionally includes the cross channels AM$\to$PM and PM$\to$AM. Across all six panels, the magnitude of the transfer function exhibits an overall increasing trend with injection ratio, indicating that under stronger injection the injected laser more readily transmits the remote source modulation.

A prominent feature, most clearly visible in the AM$\to$AM channel, is a resonance-like peak near 2~GHz, consistent with relaxation oscillation (RO) dynamics in semiconductor lasers~\cite{Lau2008JQE_FreqRespEnh_OIL,Lau2008JLT_MasterModulation_OIL_BWEnhancement}. Within the rate-equation framework, the carrier and photon (or equivalently field-amplitude) variables form a coupled second-order system around the steady-state operating point, so small perturbations can elicit a damped oscillatory response. Because this mechanism primarily reflects carrier--amplitude coupling, RO signatures are typically most pronounced in amplitude-modulated transfer channels. As the injection ratio increases from $-60$~dB to $-20$~dB, the RO-related feature shifts toward higher modulation frequencies, and the extent of this shift depends strongly on the linewidth enhancement factor. Fig.~\ref{fig:fig3}b and~\ref{fig:fig3}d visualize the AM$\to$AM transfer function as a stacked heatmap over injection ratio ($-60$ to $-20$~dB) and modulation frequency, where the vertical axis is injection ratio, the horizontal axis is $f_{\mathrm{mod}}$, and the color indicates $20\log_{10}|H|$. From Figs.~\ref{fig:fig3}b and~\ref{fig:fig3}c, when $\alpha=0$ the RO feature appears relatively smooth and its peak frequency shifts only slowly with injection ratio. In contrast, Figs.~\ref{fig:fig3}d and~\ref{fig:fig3}e show that when $\alpha=5$ the RO feature becomes more prominent and forms a clearer peak. As the injection ratio increases, the peak sharpens and shifts toward higher $f_{\mathrm{mod}}$ and can even lead to signal amplification above a certain injection ratio threshold \cite{Lau2008JLT_MasterModulation_OIL_BWEnhancement}.

This trend is summarized in Fig.~\ref{fig:fig3}f, which plots the extracted RO peak frequency versus injection ratio for $\alpha=0$ and $\alpha=5$.
For $\alpha=0$, the RO peak frequency varies only modestly, from approximately $2.0$~GHz to $2.2$~GHz across the sweep.
For $\alpha=5$, the RO peak exhibits a much larger upward shift, from approximately $2.0$~GHz up to $\sim$5~GHz over the same injection-ratio range.
Qualitatively, these observations are consistent with stronger amplitude--phase coupling introducing additional dynamical pathways and increasing the frequency selectivity of the response when $\alpha\neq 0$~\cite{Lang1982InjectionLockingProperties,HenryOlssonDutta1985LockingRangeStability}.

From a design perspective, a smaller $|H|$ indicates stronger suppression of the remote laser modulation in the injected laser output.
The RO peak shift suggests that at higher injection ratios (especially when $\alpha$ is nonzero and the RO feature becomes sharper) placing the modulation frequency sufficiently above the RO-enhanced region may be necessary to achieve stronger suppression in the injected laser output.
Conversely, at lower injection ratios the RO feature is less pronounced and its peak frequency varies less, which can provide a more stable response window at the cost of a reduced locking margin against detuning drift, as will be discussed in the following section.

\subsection{Detuning-Resolved Transfer-Function Heatmaps}

We next sweep the frequency detuning between the remote laser and the injected laser to evaluate how the above transfer functions evolve under \emph{modulated} injection and how they change across the locking boundary. We use the same rate-equation framework (Lang--Kobayashi-style formulation~\cite{Fragkos2012JLT_AmpLimiter_OIL}) and AM/PM transfer-function extraction procedure defined previously. The detuning is defined as
\begin{equation}
\Delta\nu = \nu_{\text{remote}} - \nu_{\text{injected}}.
\end{equation}

\begin{figure}[H]
  \centering
  \includegraphics[width=\linewidth]{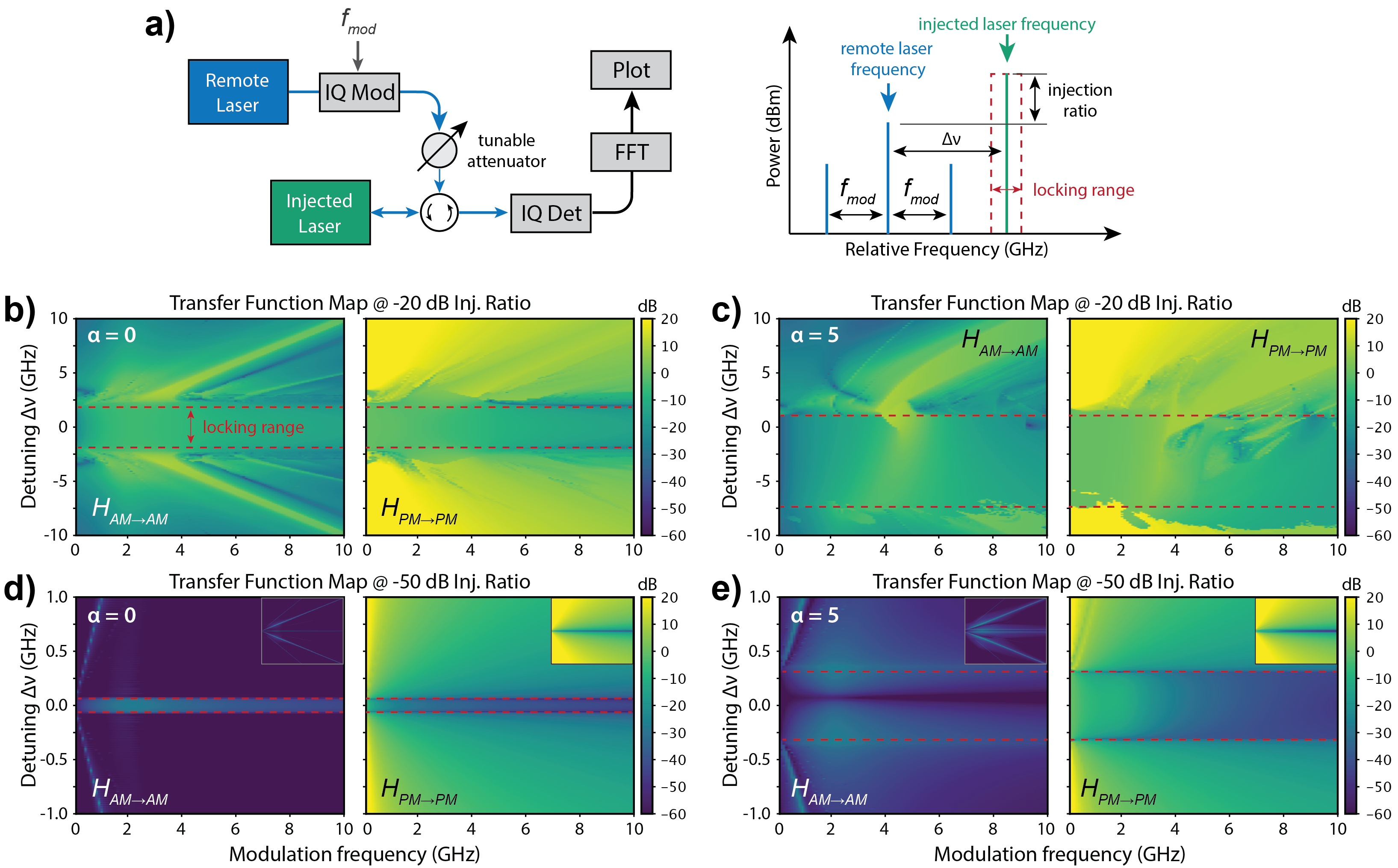}
  \caption{\textbf{Detuning-resolved transfer-function heatmaps under modulated injection ($m_a=25\%$ and $m_\phi=25\%$). a)} Conceptual overview of simulation model. Panels \textbf{b--c)} correspond to injection ratio $-20$~dB for $\alpha=0$ and $\alpha=5$, and panels \textbf{d--e)} correspond to injection ratio $-50$~dB for $\alpha=0$ and $\alpha=5$. Because the inferred locking window is much narrower at $-50$~dB, panels \textbf{d--e)} zoom the detuning axis to $\pm 1$~GHz; the full $\pm 10$~GHz range is shown as an inset in the upper-right corner of each panel. The red curve overlays the locking boundaries extracted from the 2D locking maps in Sec.~II: for $-20$~dB, $\alpha=0$ gives $-2.300$ to $+2.300$~GHz and $\alpha=5$ gives $-7.225$ to $+0.800$~GHz; for $-50$~dB, $\alpha=0$ gives $-0.075$ to $+0.075$~GHz and $\alpha=5$ gives $-0.375$ to $+0.375$~GHz. Overall, higher injection ratio provides a wider detuning margin but more readily exhibits stronger dynamical structure, whereas lower injection ratio yields a narrower but typically smoother and more stable operating window with reduced modulation remnants.}
  \label{fig:fig4}
\end{figure}

In Fig.~\ref{fig:fig4}a, the injected field is driven with a relatively large modulation depth ($m_a=25\%$ and $m_\phi=25\%$), and the detuning is swept for two representative injection ratios, $-20$~dB and $-50$~dB. For each detuning point, we compute $|H(f_{\text{mod}})|$ over $f_{\text{mod}}\in[0.1,10]\ \text{GHz}$ and visualize the results as detuning-resolved pseudo-color maps with modulation frequency on the horizontal axis and detuning on the vertical axis. This representation effectively stacks transfer functions at different detunings into a single 2D map, which facilitates identifying: (i) a detuning range where the injected laser response remains bounded and repeatable, and (ii) detuning-dependent dynamical features such as RO-related enhancement and increased sensitivity to modulation frequency~\cite{Lau2008JQE_FreqRespEnh_OIL,Lau2008JLT_MasterModulation_OIL_BWEnhancement}.

Fig.~\ref{fig:fig4}b-c show detuning sweeps at $-20$~dB for $\alpha=0$ and $\alpha=5$, respectively, while Fig.~\ref{fig:fig4}d-e show the corresponding sweeps at $-50$~dB. Overall, the larger injection ratio ($-20$~dB) corresponds to a broader detuning interval over which the injected laser response remains coherent with the remote laser, indicating a larger effective locking margin. However, within this broader interval, the transfer maps more readily exhibit pronounced dynamical structure (e.g., stronger frequency selective enhancement, localized fluctuations, or irregular regions). When $\alpha\neq 0$, amplitude--phase coupling can further amplify crosstalk between channels (AM$\to$PM and PM$\to$AM), making the operating behavior more sensitive to detuning and modulation frequency. In contrast, the lower injection ratio ($-50$~dB) yields a narrower detuning interval with reduced discontinuity across boundaries and frequencies, thereby providing a more predictable operating region with weaker remnants of remote modulation and a more controllable injected laser output. Notably, under the relatively large modulation depth used here, the low-injection case also tends to exhibit better stability than the high-injection case, in the sense that it is less prone to pronounced dynamical fluctuations or irregular structures within the inferred locking window.

From a system-design perspective, these observations highlight a trade-off: a larger injection ratio can increase the detuning margin, but may also introduce stronger dynamical effects and more pronounced AM/PM mixing, whereas a smaller injection ratio reduces the locking margin but can provide a more predictable operating window where the injected laser is still carrier-locked but experiences reduced nonlinear effects due to modulation.
This motivates the subsequent sweeps in the low injection ratio regime which further quantify how modulation depth affects the injected laser's purity and stability.

\subsection{Modulation-Depth Sweep: Robustness at Low Injection Ratio}

Based on the preceding sweeps, a consistent trend emerges: while stronger injection can provide a larger locking margin, it also introduces more pronounced nonlinear features and frequency sensitivity (e.g., sharper RO-related structure and increased dependence on modulation conditions)~\cite{LiuSlavik2020OILTutorial,Lau2008JQE_FreqRespEnh_OIL,Lau2008JLT_MasterModulation_OIL_BWEnhancement,WieczorekKrauskopfLenstra1999Bifurcations}. From a design point of view, this motivates using the lowest injection ratio that still maintains carrier-level locking and sufficient detuning tolerance. Motivated by these observations, we fix the injection ratio at $-50$~dB and the detuning at $\Delta\nu=0$, and sweep the modulation depth from $1\%$ to $100\%$ to assess robustness under strong modulation. For each modulation depth, we compute the transfer magnitude $|H(f_{\mathrm{mod}})|$ over $f_{\mathrm{mod}}\in[0.1,10]$~GHz using the same AM/PM transfer-function extraction procedure defined earlier, and plot the resulting frequency responses for the relevant channels.

\begin{figure}[H]
  \centering
  \includegraphics[width=\linewidth]{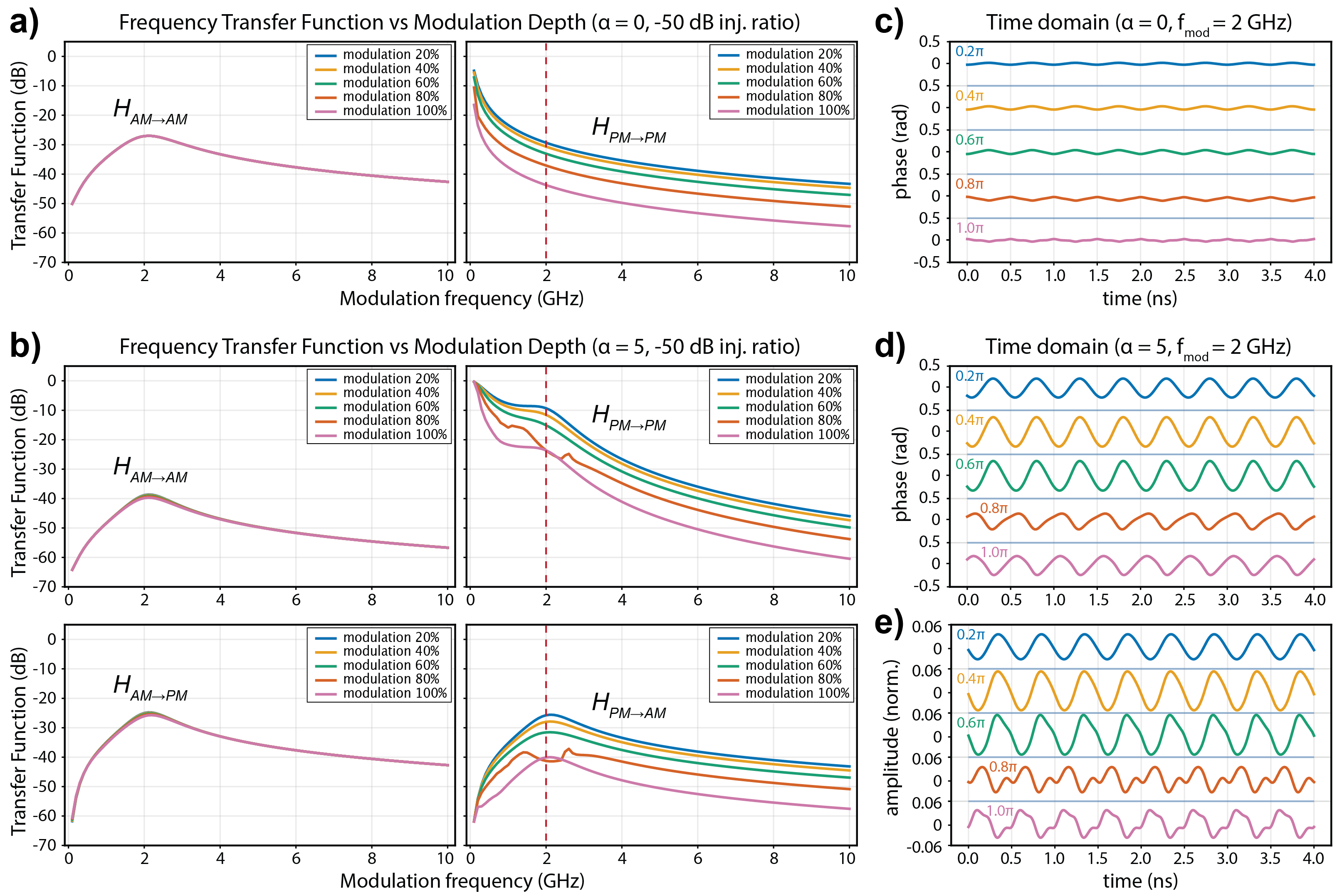}
  \caption{\textbf{Modulation-depth sweep at injection ratio $-50$~dB and $\Delta\nu=0$. a)} $\alpha=0$: AM$\to$AM and PM$\to$PM transfer functions for multiple modulation depths. \textbf{b)} $\alpha=5$: AM$\to$AM, PM$\to$PM, and PM-driven cross-channel responses, where localized distortion can appear near the RO neighborhood at high modulation depth (red dashed guide). \textbf{c)} $\alpha=0$, $f_{\mathrm{mod}}=2$~GHz: PM$\to$PM phase waveform in time domain for multiple depths. \textbf{d)} $\alpha=5$, $f_{\mathrm{mod}}=2$~GHz: PM$\to$PM phase waveform showing distortion at high PM depth (notably $\sim$80\%). \textbf{e)} $\alpha=5$, $f_{\mathrm{mod}}=2$~GHz: PM$\to$AM time-domain waveform showing a similar high-depth distortion trend.}
  \label{fig:fig5}
\end{figure}

Fig.~\ref{fig:fig5} summarizes the results under low injection.
Panels (a) and (b) show the AM$\to$AM and PM$\to$PM transfer functions for $\alpha=0$ and $\alpha=5$, respectively, for multiple modulation depths.
A clear observation is that the AM$\to$AM response is largely insensitive to modulation depth: the curves nearly overlap across depths for both $\alpha=0$ and $\alpha=5$, suggesting that under low injection the amplitude-related transfer remains highly consistent over a wide range of modulation depths.
In contrast, the PM$\to$PM channel exhibits a more systematic depth dependence: as modulation depth increases, the transfer magnitude decreases gradually, forming a distinct separation between curves across the band.

Beyond these global trends, distortion in the transfer functions becomes visible near the RO-enhanced region when $\alpha=5$ and the PM depth is large (most noticeably for the highest-depth traces, e.g., $\gtrsim 80\%$). In Fig.~\ref{fig:fig5}b, the PM$\to$PM curves develop notable non-monotonic behavior around the RO frequency (marked by the red dashed guide), deviating from an otherwise smooth roll-off. Similar behavior is also observed in the PM$\to$AM crosstalk. Importantly, this complexity remains confined near the RO frequency, while the response away from the RO peak remains comparatively smooth, and we do not observe broad-band irregularities spanning the full modulation-frequency band.

The time-domain traces at the RO frequency further corroborate the frequency-domain behavior. Fig.~\ref{fig:fig5}c shows the PM-driven phase response of the injected laser (i.e., PM$\to$PM response) at $f_{\mathrm{mod}}=2$~GHz for $\alpha=0$. Here, we see waveforms across modulation depths remain close to sinusoidal with modest variation. In contrast, Figs.~\ref{fig:fig5}d and~\ref{fig:fig5}e show the corresponding phase and amplitude response of the injected laser for the $\alpha=5$ case (PM$\to$PM and PM$\to$AM, respectively). At high modulation depth (notably around $80\%$), the waveforms exhibit clear distortion relative to a pure sinusoid, with similar (though typically milder) distortion also visible at $100\%$ depth. This time-domain distortion is consistent with the features observed in the transfer-function plots near the RO frequency. In this same modulation-depth regime near the RO peak, we additionally observe signatures of a frequency-doubling bifurcation (see time-domain plots in Fig.~\ref{fig:fig5}e), manifested as a strong enhancement of the spectral component at $2f_{\mathrm{mod}}$ in the frequency domain.

These observations are consistent with a stronger nonlinearity in the phase-driven response for large $m_\phi$, where higher-order mixing is more readily excited near the RO frequency. The system exhibits enhanced sensitivity in this frequency region and such nonlinear effects can be further amplified. Despite these nonlinear effects, the results in Fig.~\ref{fig:fig5} suggest that operating at low injection ratios can still provide comparatively stable and predictable injection locking and sideband filtering over a wide range of modulation depths as the transfer function remains low for higher-frequency components.

\section{Distributed Photoelectric Multiply-Accumulate Operations Under Optical Injection Locking}

\begin{figure}[H]
  \centering
  \includegraphics[width=0.9\linewidth]{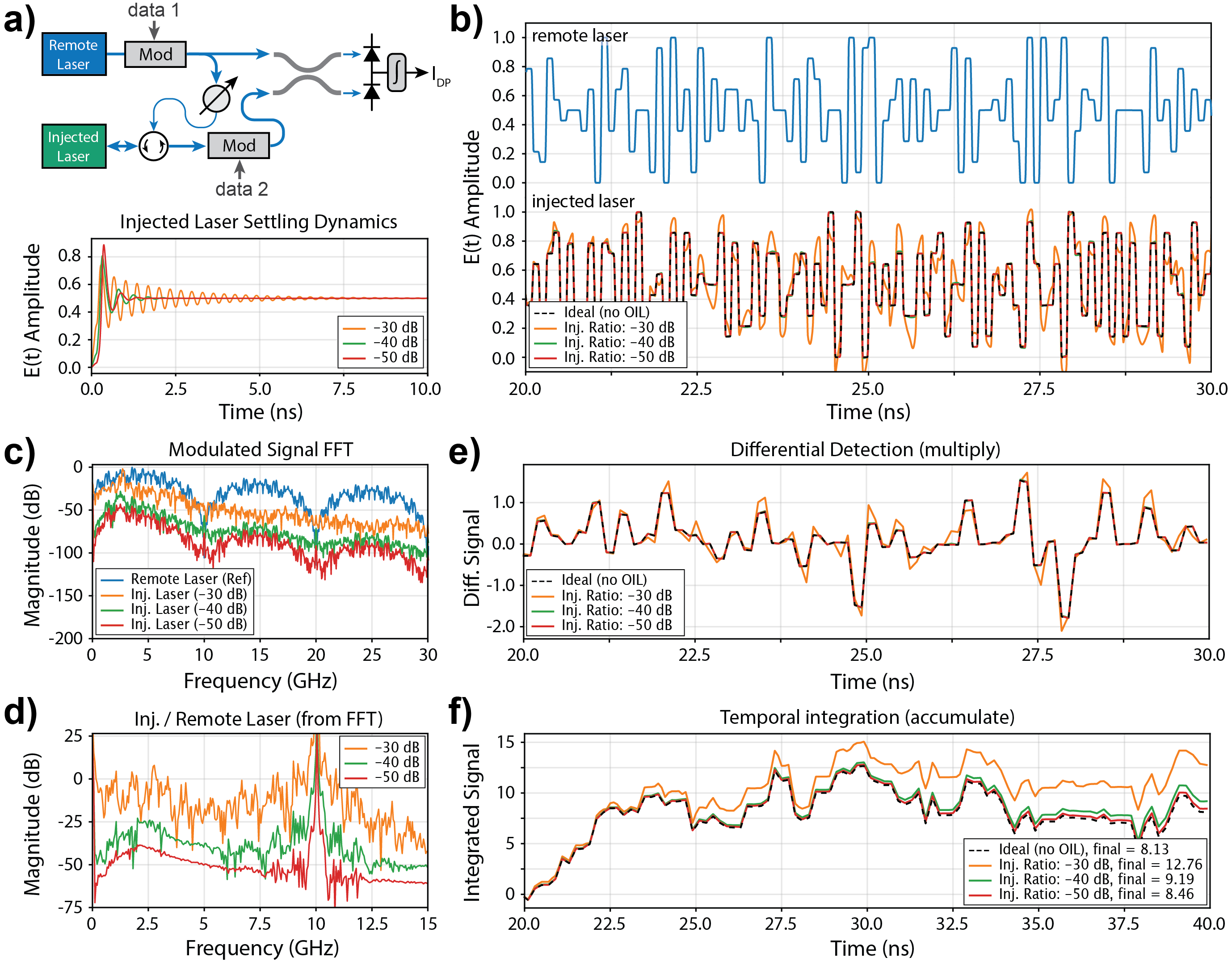}
  \caption{\textbf{Numerical evaluation of distributed coherent processing through photoelectric multiply-accumulate operations. a)} Conceptual illustration of simulated system used to evaluate distributed, coherent optical vector-vector multiplication. Plot shows injected laser field settling dynamics after start of OIL with a CW signal (0--10~ns, unmodulated). \textbf{b)} Modulated remote laser (top) and modulated injected laser fields (bottom) during OIL for injection ratios of $-30$, $-40$, and $-50$~dB. Black dashed line compares the ideal case with no injection. \textbf{c)} FFT of the remote and injected laser fields from \textbf{b)}, showing reduced modulation-related spectral components at lower injection ratios. \textbf{d)} Ratio of the FFT results in \textbf{c)} for frequencies between 0 and 15 GHz. The extracted transfer function from the broadband modulation in \textbf{b)} exhibits an RO peak near $\sim$2~GHz and lower residual modulation at lower injection ratios. \textbf{e)} Modeled BPD differential output from interference and differential detection. \textbf{f)} Temporal accumulation of the signal in \textbf{e)} where lower injection ratios yield results closer to the ideal baseline.}
  \label{fig:fig6}
\end{figure}

We finally evaluate injection locking under real symbol modulation, moving beyond single-tone transfer-function sweeps to modulation formats relevant for optical processing applications. Using the same rate-equation framework, we modulate the remote laser and the injected laser with symbol sequences and numerically evaluate their interference using an ideal homodyne detection model. This provides a direct time-domain view of symbol-level photoelectric multiplication enabled by coherent interference~\cite{RahimiKari2024Optica_CoherentPlatform,Hamerly2019PRX_PhotoelectricMultiplication,Chen2023NatPhoton_VCSEL,Zhou2025_QA_PhotonicHomodyne}. Fig.~\ref{fig:fig6}a illustrates the simulated system where a remote laser modulated with a time-varying analog signal is used to injection lock the local (injected) laser. The local laser (now a copy of the unmodulated remote laser) is then modulated with a different time-varying analog signal before the two optical paths are interfered in a 3 dB coupler and detected using a simulated balanced photodetection scheme. The plot in Fig.~\ref{fig:fig6}a shows the remote laser field amplitude in the time domain over the first 10~ns immediately after OIL starts. Because the injected laser requires a finite settling interval after injection is applied, we conservatively wait until after 20~ns to begin symbol modulation. This temporal delay reduces ambiguity between locking transients and symbol-driven dynamics. For these simulations, we assume a fixed phase relationship between the remote and injected lasers which maximizes the interference signal (i.e., $\Delta\phi = \pi/2$). We note that after OIL is established, this settling time is not needed between sequential processing operations.

To evaluate symbol-level coherent readout under injection locking, we use a real-valued modulation format in which the discrete symbol stream is arranged as antipodal symbol pairs. For each randomly selected symbol magnitude $a$, two consecutive symbols are assigned as $+a$ and $-a$. To encode $-a$, a $\pi$ phase shift is applied to the signal during the duration of the negative symbol (not shown in Fig.~\ref{fig:fig6}a). This scheme shares the DC-balanced property of multilevel Manchester-type encoding~\cite{Chung2012EMC_PAMn}, while differing from conventional extended Manchester code formulations in that the sign inversion is implemented across consecutive symbols rather than as a mid-symbol transition within a single symbol period. The main purpose of this pairing is to suppress the low-frequency and DC content of the modulation waveform. Since each adjacent symbol pair has equal magnitude and opposite sign, its average is zero which eliminates changes in the DC optical power as seen by the injected laser (i.e., the signal maintains a constant injection ratio).

After symbol generation at 5 Gbaud, the discrete sequences are converted into continuous-time drive signals by zero-order hold and then filtered by a 20~GHz low-pass filter, producing band-limited real-symbol waveforms for both paths. In the remote branch (Path~A), the filtered waveform is applied on top of a DC optical carrier. In the injected branch (Path~B), the injected-laser field is first obtained from the rate-equation dynamics, and the filtered Path~B waveform is then added to this field. To isolate the useful symbol-driven response from initial locking transients, the first 200 symbols are set to zero, corresponding to a pure DC segment from 0 to 20~ns. The following 200 symbols, spanning 20 to 40~ns, form the active modulation region. This arrangement allows the injected laser to settle before symbol modulation is applied, so that the subsequent time-domain interference and modeled balanced-detection results can be interpreted primarily as symbol-driven behavior rather than startup dynamics. We note that while our symbol rate is relatively modest for this analysis, much higher rates are further from the RO frequency and are expected to be less affected by the nonlinearities observed in Fig.~\ref{fig:fig5}.

Fig.~\ref{fig:fig6}b plots both the remote laser after modulation begins (top) and the injected laser field (bottom) for injection ratios of $-30$, $-40$, and $-50$~dB after OIL and subsequent modulation is applied. We additionally include the ideal case as a reference (dashed black line) where the local laser is modulated without injection from the remote laser. We observe from the response of the injected laser that during symbol modulation (after 20~ns), stronger injection ratios exhibit more visible residual oscillatory components and larger waveform distortion relative to the ideal case, consistent with increased coupling of source modulation into the injected laser output.
Among the tested conditions, the $-50$~dB case shows the smallest distortion and most closely approaches the ideal baseline, as expected.

To quantify the frequency-dependent coupling of the remote laser modulation into the injected laser output as a function of frequency, we compute the FFT of the remote and injected laser field during modulation from the remote laser, but before modulation of the injected laser's output (i.e., after the circulator in Fig.~\ref{fig:fig6}a). The results are shown in Fig.~\ref{fig:fig6}c where lower injection ratios correspond to weaker spectral components at the injected laser's output, indicating reduced coupling of the modulated signal from the remote laser after OIL.
Fig.~\ref{fig:fig6}d reports the corresponding ratio of spectral power between the injected and remote lasers, which follows a trend consistent with the small-signal transfer analysis in Fig.~\ref{fig:fig3}e. Here, we observe an RO-related enhancement appears near $\sim$2~GHz, followed by an overall roll-off, and lower injection ratios yield lower residual modulation across frequency.

Fig.~\ref{fig:fig6}e shows the modeled balanced-photodetector (BPD) differential output obtained by interfering the remote-modulated signal and the locally modulated injected laser signal.
Under appropriate normalization, the differential output is proportional to the interference cross term $2E_1E_2$, producing positive values when the symbol polarities match and negative values when they differ.
Fig.~\ref{fig:fig6}f plots the modeled temporal accumulation (integration) of the BPD differential signal.
As the injection ratio decreases, the integrated result approaches the no-injection baseline more closely over the simulated window, indicating improved dot-product accuracy and reduced contamination from residual remote modulation coupling.
Overall, these symbol-level results reinforce the design guideline suggested by the preceding sweeps: when carrier-level locking can be maintained, using a lower injection ratio is more likely to yield a ``purer'' injected laser output with reduced remote modulation remnants, improving interference-based computation fidelity.

\section{Conclusion}

This work presents a theoretical and simulation-based assessment of optical injection locking as a mechanism for maintaining mutual coherence between a remotely modulated source and a local processing node, enabling distributed, real-time coherent photonic computing without optical-to-electrical conversion. Using a semiconductor-laser rate-equation framework, we systematically explore how injection ratio, frequency detuning, linewidth enhancement factor $\alpha$, and modulation conditions jointly determine the viability of this approach.

Three main findings emerge. First, 2D locking-region maps confirm that stronger injection broadens the carrier-level locking interval but, for $\alpha \neq 0$, introduces increasing asymmetry and localized dynamical instabilities near the locking boundary. Second, frequency-resolved AM/PM transfer-function analysis spanning injection ratio, detuning, and modulation-depth sweeps reveals a fundamental design trade-off: higher injection ratios widen the detuning tolerance but also strengthen the coupling of remote modulation into the injected laser output, particularly near the relaxation-oscillation frequency where resonant enhancement and amplitude–phase crosstalk are most pronounced. Conversely, lower injection ratios produce a narrower but smoother operating window in which remote modulation remnants are strongly suppressed across the modulation band. Third, end-to-end symbol-sequence simulations with balanced detection and temporal integration corroborate these trends at the system level. These results indicate that reducing the injection ratio yields integrated dot-product readouts that converge toward the ideal no-injection baseline, confirming improved computational fidelity.

Together, these results point to a clear design guideline in which the lowest injection ratio should still provide adequate carrier-level locking margin and detuning tolerance for the target deployment environment. In this regime, the injected laser acts as a carrier-locked yet modulation-isolated local oscillator, maximizing the purity of the coherent reference available for interference-based computation. Modulation-frequency placement should also account for $\alpha$-dependent RO dynamics, avoiding the RO-enhanced spectral region where residual transfer and nonlinear mixing are strongest.

Several directions remain for future investigation. Incorporating spontaneous-emission noise and realistic fiber-channel impairments (e.g., polarization drift, chromatic dispersion, and amplified spontaneous emission from in-line EDFAs) will be essential for translating these deterministic maps into practical operating margins. Experimental validation is needed to confirm the predicted trends under real device variability. Beyond validation, extending the framework to wavelength-division-multiplexed carriers, higher-order modulation formats, and full matrix-vector-multiply architectures will help bridge the gap between the single-channel analysis presented here and system-level throughput and energy-efficiency projections for distributed coherent photonic accelerators.

\section{Methods}
\subsection{Defining the Locking Range}
After plotting the parameter maps of $\Delta I/\bar{I}$ and $\Delta\phi$, a fairly clear transition can be observed in both cases: one side corresponds to a low-value region with comparatively steady behavior, whereas the other corresponds to a region with increased residual oscillation or phase drift. To better identify the approximate range of this transition, we tested a series of threshold values and found that the boundary in the present simulations lies roughly near 0.1 for the intensity metric and 0.1~rad for the phase metric. Accordingly, we use $\Delta I/\bar{I}\le 0.10$ and $\Delta\phi\le 0.10~\mathrm{rad}$ as empirical operational thresholds to help delineate the boundary of the locking region. For the intensity metric, $\Delta I/\bar{I}\le 0.10$ indicates residual variation below approximately 10\% of the mean intensity, which in this work generally separates approximately steady trajectories from states with more pronounced oscillation. For the phase metric, $\Delta\phi\le 0.10~\mathrm{rad}$ similarly corresponds to a small residual phase excursion over the analysis window, indicating that the field phase remains approximately stationary. These numerical values are therefore not intended as strict theoretical criteria for injection locking, but rather as empirical thresholds used to more clearly identify and outline the locking-region boundary in the present simulations. This distinction is relevant because prior studies have shown that optically injected lasers can exhibit oscillatory states with bounded phase and average-frequency entrainment near Hopf-mediated locking transitions, even when the intensity is no longer strictly steady-state~\cite{Kelleher2012PRE_BoundedPhase,Pausch2012NJP_QDInjectionDynamics}. In that sense, the present thresholds are used to define a conservative practical operating window emphasizing jointly stable amplitude and phase behavior, rather than the broadest possible notion of frequency locking.

\vfill

\end{document}